# J/Psi suppression, percolation model and the critical energy density in AA collisions at SPS and RHIC energies with the account of centrality.

G. Feofilov[1†] and O. Kochebina[2]

*V.A.Fock Institute of Physics of St-Petersburg State University*

† E-mail: [1]feofilov@hiex.niif.spb.su, [2]kochebina@gmail.com

**Abstract**

Experimental data on the onset of a so-called "anomalous" J/psi suppression in PbPb collisions at the SPS are considered in the framework of string percolation model. This onset is observed at a certain collision centrality, characterized by the number of nucleons – participants (Npart) related to the impact-parameter. Modified Bjorken formula calculations are performed for the local energy densities in AA collisions at different impact-parameters and with detailed account of the latest data available on the charged particles densities at midrapidity. Finally we compare variations of mean local energy and of string densities and match the occurrence of the critical percolation phenomenon with the critical energy density value, considering them at the same values of Npart. Similar analysis was performed for recent RHIC data and the results are discussed.

**Introduction**

The phenomenon of "anomalous" suppression of J/psi (Fig. 1), that was observed in nuclear collisions by the experiments Na38, Na50, Na60 at CERN SPS (see [1] and reference there in), can indicate the onset of deconfinement transition predicted in the finite temperature QCD[2,3]. Onset of quarkonia melting above a certain temperature / energy density threshold should provide an "unambiguous" signature of quark-gluon plasma formation [3]. Recent experimental data [4] and theoretical discussions (see overview in[5]) show that some problems are still to be solved, the magnitude of suppression and its behavior with the collision energy and in various rapidity regions are still not well understood [5].

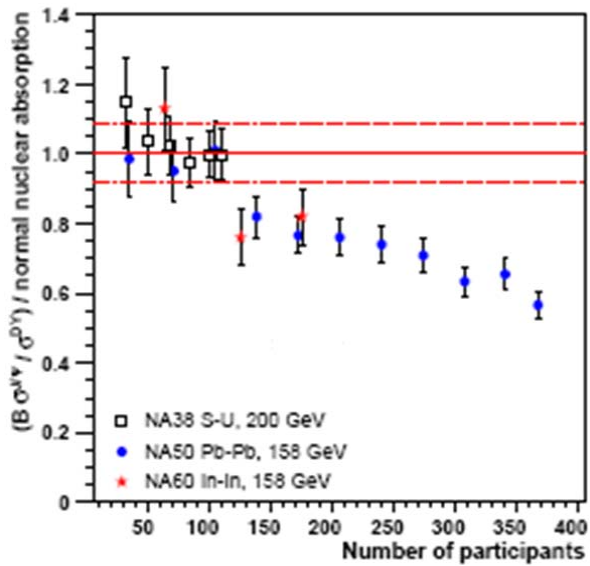

Fig. 1. Experimental data for yield of J/psi in nuclear collisions at the SPS . (see [1] and reference there in)

The onset of the mentiond "anomalous" J/psi suppression in PbPb and InIn collisions at the SPS is observed at a certain collision centrality [1]. The last one is characterized by the number of nucleons – participants (Npart) related to the impact-parameter. So, one can apply the well established "critical" values of the Npart, marking the onset of some new phenomena or processes reached at certain collision energy density, for analysis of various approaches. In the present study the experimental data on the onset of J/psi suppression are considered in the framework of string percolation model [6]. A geometrical effect, due to clustering and percolation of strings, may be considered as a condition that is necessary in the pre-equilibrium stage of nuclear collision to achieve deconfinement and subsequent quark-gluon plasma formation [7].

The results of percolation model are compared with the local energy density behavior obtained using the calculations based on the Bjorken formula [8]. Modifications of the last one were done in order to provide the estimates for the mean local energy densities in AA collisions relevant to the different impact-parameters. The detailed account of the latest data on the charged particles densities at midrapidity reached at various nuclear collision centralities is also applied. Similar analysis is presented for the recent RHIC data and the results are discussed below.

**Estimates of local energy density for AA collisions at SPS CERN (√s =17,3 GeV) as a function of centrality.**

We start first with the study in the framework of the modified Bjorken formula of centrality dependence of the local energy density reached in nucleus-nucleus collisions. Centrality of nuclear collisions is described usually by the impact-parameter $b$, representing the distance between the centers of two colliding nuclei. It is well known that this parameter $b$ in any theoretical approach could be almost directly connected with the experimentally observed quantity – the number of nucleons - participants in nuclear collision (Npart).

Classical Bjorken formula [8] deals with the central ($b=0$) collisions of nuclei with regular spherical shape. This central collision of two identical nuclei with mass number $A$, with the Lorentz-factor $\gamma$ ($\gamma = E/m \gg 1$, where $m$ - nucleon mass, $E$ - the energy per nucleon) could be represented as a collision of two Lorentz compressed disks. At the time $t = 0$ nucleons of colliding nuclei interact each other and the hadronic fireball is formed [8]. During the subsequent moments of time the fireball extends with speed on the order of the speed of light. The energy density is equal to the ratio of total energy of the secondary particles which are outgoing from interaction area in the central rapidity region and the initial collision volume.

The energy density is given by the classical Bjorken formula [8]:

$$\varepsilon = \frac{N_{part}}{S}\frac{d\langle E\rangle}{dy}\frac{1}{2\tau} = \frac{N_{part}}{S}\frac{3}{2}\frac{dN_{ch}}{dy}\langle E\rangle\frac{1}{\tau} \quad (1),$$

where Npart – the number of nucleons-participant, $\frac{d\langle E\rangle}{dy}$ - energy of secondary particle per rapidity unit, <E> ≈400 MeV – is the mean energy per secondary particle in the central rapidity region, $\frac{dN_{ch}}{dy}$ - charged particle density at midrapidity, $\tau$ – transit "proper time", S – overlap area of colliding nuclei. The collision volume is defined by the area S and time $\tau$. This area S is the transverse area of the fireball that is usually taken as $S = \pi R^2$ for the central collisions, where $R = (1,2 A^{1/3} fm)$ is the radius of nucleus. Time $\tau$ is taken as $\tau \approx 1$fm).

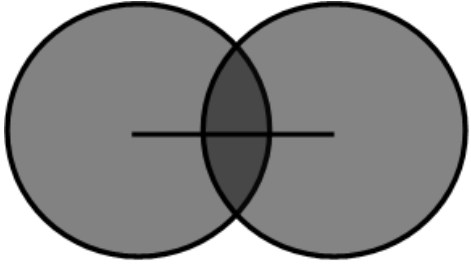

Fig.2. Determination of overlap area S as function impact parameter *b*.

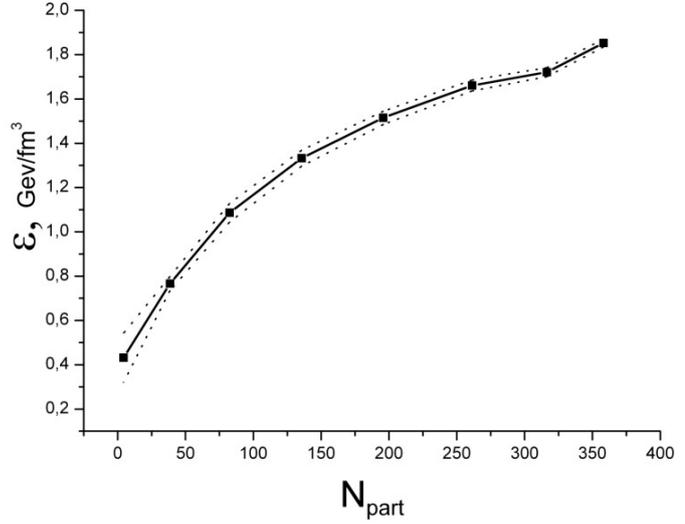

In our work here the local energy densities at AA collisions are estimated by using the Bjorken formula [8] modified for the account of centrality dependence.
So, one has the following local energy density as a function of centrality:

Fig.3. Dependence of the energy density on the number of participants for PbPb collisions at **17.3** GeV

$$\varepsilon(b) = \frac{N_{part}(b)}{S(b)} \frac{3}{2} \frac{dN_{ch}}{dy}(b) \langle E \rangle \frac{1}{\tau} \quad (2)$$

Our estimates are based on consideration of the mean local energy density, defined for a certain overlap area, as a function of centrality for collision of nuclei with diffuse edges (using the Woods-Saxon distribution). This means that the interaction volume, provided by the certain number of Npart, is estimated in our case as product of S(b) and τ (time τ in our calculation is the same as in Bjorken work). As to the overlap area S (b), - we would like to note here that our approach is based on two main assumptions:

1) First of all, the pure geometrical definition of the overlap area S=S(*b*) of two nuclei characterized by the radii R is done for the given impact parameter *b* (see for example Fig. 2).
2) Then the number of participants Npart (b) is found by the simulations using the MC Glauber model [9] relevant to this concrete overlap area S. Diffuse edge effects are accounted for by using the detailed Woods-Saxon distributions. The estimates of systematic errors of values of Npart(b) due to the cuts of the tails of distributions outside the given radii R are also performed (they are shown to be within 5%).

It is also important to note that the latest available detailed experimental data [10] on $\frac{dN_{ch}}{dy}$ (b) are used. Thus, the mean local energy density dependence on participant number is received on the Fig. 3. As one can see, the energy density grows with growing number of participants and reaches the value of about $\varepsilon \approx 1.8$ GeV/fm$^3$ in the very central PbPb collisions (Npart~350) at $\sqrt{s}$ =17.3 GeV.

We have to note here that this value differs from the previous results (see, for example [11] where $\varepsilon \approx 2.4$ GeV/fm$^3$ was obtained for the very central PbPb collisions). Second note is that the critical energy density relevant to the Npart=110 could be obtained from the results of Fig. 3 to be equal approximately 1.2-1.3 GeV/fm$^3$.

**Estimates of local percolation parameter for AA collisions at SPS CERN (sqrt(s)=17,3 GeV) as a function of centrality.**

Fusion of colour strings [12,13] stretched between the valence as well as sea quarks belonging to projectile and target in the process of nucleus-nucleus collision could be an interesting new phenomenon preceding the formation of the quark-gluon plasma. Color strings may be viewed as small discs in the transverse space, $\pi r_0^2$, $r_0$=0.2-0.25 fm, filled with the color field created by the colliding partons. With growing energy and/or atomic the number of colliding particles, the number of strings grows and they start to overlap, forming clusters (Fig.4). At a critical density a macroscopic cluster appears that marks the percolation phase transition [7]. Percolation parameter is defined as:

Fig.4. Formation string. From left to right energy and/or number of participant grow [7].

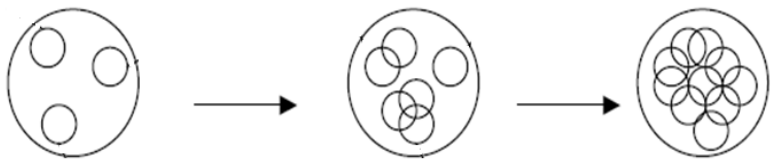

$$\eta(b) = N_{str}(b)\pi r_0^2 / S(b) \quad (3)$$

where $\pi r_0^2$- string transverse area ($r_0$=0.2-0.25 fm), $N_{str}$ – the number of strings, S – nuclear overlap area of colliding nuclei (as it has been defined early). Percolation parameter is considered as a function of $b$. Number of strings (Nstr) and the density of overlapping strings also depend on the centrality of collision. Critical value of the percolation parameter $\eta$ could be calculated from the geometrical considerations, it equals 1.15 at $r_0$=0.2 fm [14]. Dependence of $\eta$ on the Npart was calculated in the same framework as described in the previous section (see Fig.5). The assumption is used that the critical value of $\eta$=1.15 is reached at Npart=110 in nuclear collisions at $\sqrt{s}$=17.3 GeV. Thus the "critical" string number Nstr could be also defined straightforwardly.

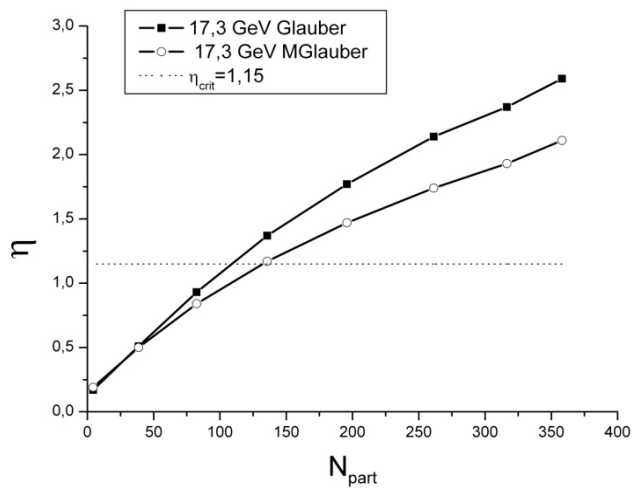

Fig.5. Percolation parameter $\eta$ as a function of number of participants for PbPb collisions at $\sqrt{s}$ =17,3 GeV)

(We obtain the values of Nstr=500 in case of $r_0$=0.2 fm and Nstr=322 for $r_0$=0.25 fm, both are relevant to the Npart=110 in PbPb collisions at $\sqrt{s}$=17.3 GeV.). Calculations presented on the Fig.5 were done for 2 cases: standard Glauber model and the modified one (MGM or MGlauber) [9] where effects of nucleon participant stopping are taken into account). Estimates of number of nucleon-nucleon collisions (Ncoll) are different (lower) in the framework of the MGM, thus the lower string density is produced in the last case (see Fig.5). (Number of Nstr is defined as Nstr=Nv+Nsea, where we may assume that the number of valent strings Nv=Npart and the number of strings produced by the sea quarks is defined at the given collision energy by the number of nucleon nucleon collisions: Nsea=2Ncoll). The MGM is taking into account the nucleon stopping, therefore the the number of collisions is less then in a standard Glauber ,and, therefore, the onset of the percolation critical η is shifted to the larger numbers of Npart shown on the Fig. 5.

**Comparison of energy density and percolation parameter values for different impact-parameters.**

Finally we compare the energy density and percolation parameter obtained at the same fixed number of participants. Calculations are shown on the Fig.6 again for cases of standard Glauber model and for the MGM [9]. We conclude here:

(1) In nuclear collisions at √s = 17. 3 GeV the onset of the percolation phase transition could be in line with hypothesis of phase transition between hadrons gas and quark - gluon plasma at center-of-mass system energy achieved at $N_{part}$ = 110;
(2) Energy density corresponding to this transition should be equal approximately 1.2-1.3 GeV/fm$^3$.

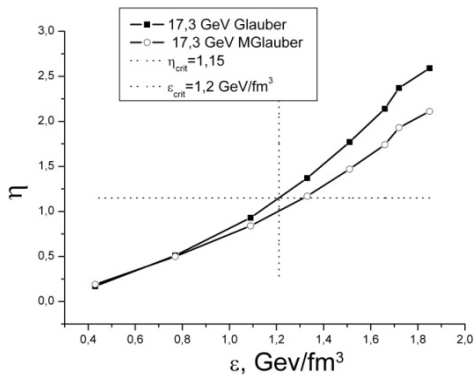

Fig.6. Comparison of energy density and percolation parameter values obtained for nuclear collisions at √s = 17. 3 GeV

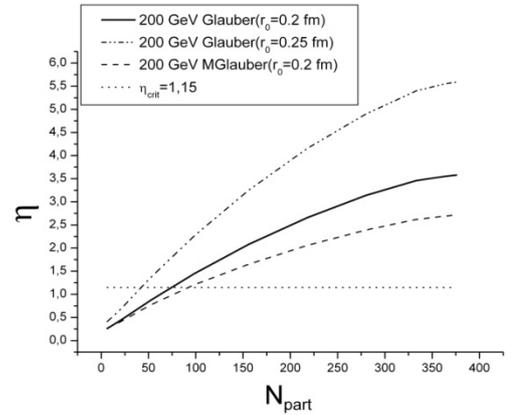

Fig .7. Percolation parameter as a function of number of participants ( AuAu, √s =200 GeV)

**RHIC (√s =200 GeV) AuAu data and critical value percolation parameter and energy density.**

Finally we made similar estimates for AuAu collisions at RHIC energy $\sqrt{s}$ =200 GeV. We estimated the percolation parameter by using formula (3) (see, Fig.6). By the dashed lines we reveal the systematic uncertainty estimates. The latter might be due to the inaccuracies of definition of number of strings, string radius and overlap area. With the growing energy the number of strings increases, and the results of the Fig.6 give, that at $\sqrt{s}$ =200 GeV the critical percolation parameter should be reached at number of participants $N_{part}$ in the region 30 to 95. At the same time the experimental data for the onset of the J/psi suppression at midrapidity in AuAu collisions at $\sqrt{s}$ =200 GeV indicate the value of the critical Npart ~ 160 [4].

**Conclusions.**

(1) There is a reasonable agreement between the hypothesis of the onset of percolation transition and the relevant critical energy density in PbPb collisions at CERN SPS ($\sqrt{s}$=17.3 GeV); The value of the critical energy density is defined to be approximately 1.2-1.3 GeV/fm$^3$.
(2) Contradiction obtained here between the theoretical calculations in the framework of string percolation model and the experimental results for J/psi suppression at RHIC energy (200 GeV) requires additional investigations.

**Acknowledgements.** There studies were partially supported by the grant RNP.2.2.2.2.1547 by the Russian Ministry of Education and Science.

**References:**

[1] B. Allesandro et al., (NA50), Eur. Phys. J.C 39, 335 (2005); NA60 Colaboration, PRL99 132302 (2007).
[2] H.Satz, Nucl.Phys. A 418, 447 C (1984).
[3]T. Matsui, H.Satz, Phys. Lett. B178, 4 (1986).
[4] PHENIX Collaboration, Phys.Rev.Let,98, 232301(2007).
[5] Y. Zhang, Overview of Charm Production at RHIC, report at QM2008 (2008);
R.G.de Cassagnac, Quarkonia Production in Cold and Hot Matters, report at Quark Matter 2008, Jaipur, February 6$^{th}$ (2008).
[6] N. Armesto, M.A. Braun, E.G. Ferreiro and C. Pajares, Phys. Rev. Lett. 77, 3736 (1996).
[7] C. Pajares // Eur.Phys.J. C43, 9-14 (2005).
[8] J.D. Bjorken, Phys.Rev.D27, 140(1983).
[9] G.Feofilov,A.Ivanov, Journal of Physics: Conference Series 5, 230–237 (2005).
[10] F. Antinori, et al.// Journal of Physics: Conference Series 5, 64–73 (2005).
[11] A. Andronic and P. Braun-Munzinger// arXiv:hep-ph/0402291v1 (2004).
[12] A.Capella, U.P.Sukhatme, C.--I.Tan and J.Tran Thanh Van, Phys. Lett. B8, 68 (1979); Phys. Rep. 236, 225(1994) .
[13]A.B.Kaidalov, Phys. Lett., 116B,459 (1982) ;A.B.Kaidalov K.A.Ter-Martirosyan, Phys. Lett., 117B, 247 (1982).
[14] J.Dias de Deus and A. Rodrigues// Phys. Rev. C 67, 064903 (2003).